# A plasmonic painter's method of color mixing for a continuous RGB palette

**Authors**

Claudio U. Hail[1], Gabriel Schnoering[1], Mehdi Damak[1], Dimos Poulikakos[1]*, Hadi Eghlidi[1]*

**Affiliations**

[1] Laboratory of Thermodynamics in Emerging Technologies, ETH Zürich, Sonneggstrasse 3, CH-8092 Zürich, Switzerland.

* Correspondence and requests for materials should be addressed to D.P. (email: dpoulikakos@ethz.ch) or H.E. (email: eghlidim@ethz.ch).

**Abstract**

**The ability of mixing colors with remarkable results had long been exclusive to the talents of master painters. By finely combining colors at different amounts on the palette intuitively, they obtain smooth gradients with any given color. Creating such smooth color variations through scattering by the structural patterning of a surface, as opposed to color pigments, has long remained a challenge. Here, we borrow from the painter's approach and demonstrate color mixing generated by an optical metasurface. We propose a single-layer plasmonic color pixel and a method for nanophotonic structural color mixing based on the additive RGB color model. The color pixels consist of plasmonic nanorod arrays that generate vivid primary colors and enable independent control of color brightness without affecting chromaticity, by simply varying geometric in-plane parameters. By interleaving different nanorod arrays, we combine up to three primary colors on a single pixel. Based on this, two and three color mixing is demonstrated, enabling the continuous coverage of a plasmonic RGB color gamut and yielding a palette with a virtually unlimited number of colors. With this multi-resonant color pixel, we show the photorealistic printing**

**of color and monochrome images at the nanoscale, with ultra-smooth transitions in color and brightness. Our color mixing approach can be applied to a broad range of scatterer designs and materials, and has the potential to be used for multi-wavelength color filters and dynamic photorealistic displays.**

## Introduction

Materials that induce vivid colors through their structuring, manifest themselves spectacularly in nature, ranging from the wings of butterflies and the wing cases of beetles to opal gemstones[1]. Artificially generating colors in a similar manner requires careful tailoring of a surface structure at the scale of the wavelength of visible light or smaller. Metal nanoparticles, due to their strong scattering of light and large spectral tunability, offer a promising platform for inducing structural colors[2,3]. As opposed to dye and pigment-based colors, such plasmonically induced colors are less susceptible to color fading and enable the representation of colors with diffraction limited resolution[4]. These optically resonant nanostructures can be tailored to exhibit intense colors when illuminated with white light. Depending on the structure geometry, the resulting colors are observed in reflection[5–11] or in transmission[12–15], and are generated using subtractive[7–9] or additive[5,10] color schemes. In this respect, the scattering of individual nano-optical structures such as rods[5], disks[6,10,16], holes[14] or metal-insulator-metal structures[9,11,17] have been explored to generate colors with maximum saturation and high resolution. In several other notable studies, dielectric scatterers exhibiting Mie resonances have also been employed for generating vivid colors[18–21].

While with these structures the color saturation is limited by the resonance quality of the individual optical scatterer, the combination of individual resonances with lattice modes enables resonance sharpening and thus more vivid colors[10,22–24]. In particular, spectrally overlapping the particle scattering with Wood's anomalies of the lattice creates sharp spectral features[10,23]. However, to extend the available color gamut, methods for the control of the luminance of highly saturated individual colors, as well as their precise mixing are required. Color mixing can be obtained by dividing each pixel into many spatially separated sub-pixels. For example, by selecting five spatially separated sub-pixels which were colored in either black, white or in the three primary colors, the variation of color saturation and lightness was demonstrated[25]. Another approach is to vary the concentration of two different scatterers each exhibiting a different color in a square or hexagonal lattice[16,19]. Although extending the color palette, both approaches introduce spatial non-uniformities and are limited to specific mixing ratios, which limits the total number of colors that can be represented. The representation of photorealistic images with smooth transitions between different color hues and tones thus demands more flexibility in color mixing and in setting the brightness of individual colors.

Here, we demonstrate a plasmonic color pixel and color mixing approach with a new level of control over color luminance and chromaticity. We attain seamless mixing of colors at the nanoscale, which enables the realization of countless numbers of color variations, and as a consequence, a continuous coverage of the color gamut. Our color pixels consist of silver nanorods arranged in a periodic rectangular lattice on a transparent glass substrate. By spectrally overlapping the scattering resonance of the silver nanoparticles with the first Wood's anomaly of the lattice, we obtain vivid plasmonic color primaries that cover 55% of the sRGB color gamut. As a first requirement for color mixing, we show that these single color pixels allow setting the color luminance without affecting the chromaticity of the color. Interestingly, seamless variations in colors are obtained by creating a multi-resonant dual or triple color pixel consisting of a superposition of two or three different colors and varying their corresponding luminance. This enables the representation of smooth photorealistic color and brightness transitions and a continuous coverage of over 39% of the sRGB color gamut. This is a large step forward with respect to the state of the art, where, instead, discrete color hues and tones were reproduced[4,5,18–21,26,6,7,9–11,13–15].

## Results & Discussion

Vivid primary colors are generated in reflection by taking advantage of the strong scattering of subwavelength sized silver nanorods with length $L$ and width $W$. As shown in Fig. 1a, a single color pixel consists of a transparent substrate and a rectangular lattice of nanorods, formed by a 2-dimensional nanorod array with periodicities $P_x$ and $P_y$ along the $x$ and $y$ direction respectively, where the nanorod long axis extends along the $y$ direction. The plasmon resonance of an individual nanorod is set by the nanorod width and length, and is spectrally broad due to dissipation in the metal[27,28]. At a wavelength $\lambda$ the diffractive coupling of nanorods is maximized by setting the lattice periodicity to $P_x = \lambda/n$, which corresponds to the condition of the first Wood's anomaly of the lattice in the substrate of refractive index $n$. At this wavelength, the first diffraction orders in the substrate propagate at a grazing angle and greatly enhance the scattering of the nanorods[29]. Overlapping the broad scattering resonance of the nanorods with the narrow Wood's anomaly resonance, creates a sharp Fano resonance[30]. This effect has been studied theoretically and experimentally for obtaining high-quality resonances from metallic nanoparticle arrays[23,31–33]. In this respect, the correct combination of the length and width of the nanorod and the lattice periodicity $P_x$, enables the generation of vivid primary colors. Depending on the material of the scatterers, the particle and lattice resonance can be overlapped and finely tailored to give narrow spectral features over the entire visible range. By illumination with white light polarized along the $y$ direction at normal incidence, vivid colors are obtained in reflection. The reflected intensity and spectral narrowing due to the lattice arrangement is typically dependent on the number of periodicities reproduced. Here, we choose a minimum of 10 periods in *x* direction for the largest $P_x$ value, which results in a smallest pixel size of 4.26 μm. A smaller unit cell size may be used at the expense of spectral broadening (see Supporting Information S1 for an analysis of the unit cell size dependence).

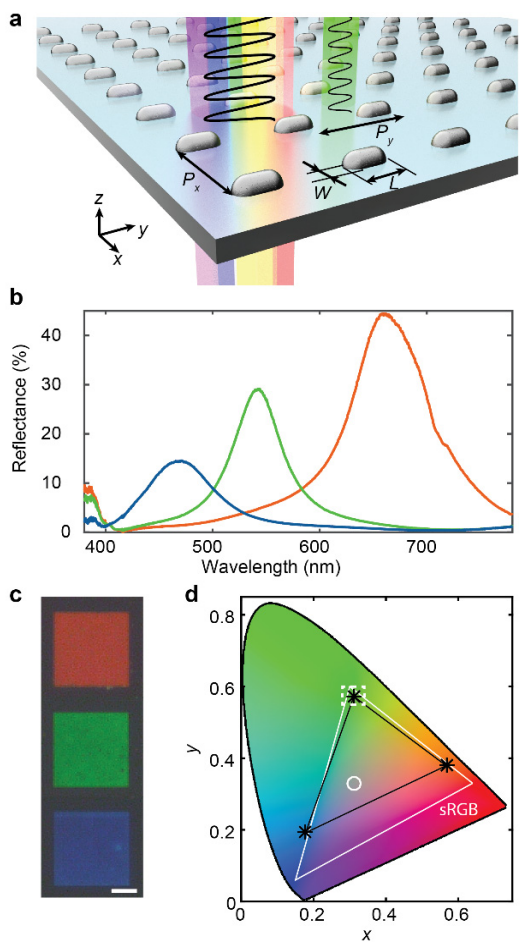

**Figure 1. | Design and performance of the plasmonic RGB color primaries.** (a) Schematic of the single color plasmonic pixel consisting of a lattice of silver nanorods on a glass substrate. White light illumination polarized along the long axis of the nanorods results in distinct colors observed in reflection. The length *L* and width *W* of the nanorods set the local surface plasmon resonance, the periodicity along the *x* direction,

$P_x$, sets the lattice resonance and the periodicity along the $y$ direction, $P_y$, sets the color luminance. (b) Measured reflectance spectra of the red, green and blue plasmonic primary colors. The respective nanorod lengths and widths are $L$ = 143, 102, 63 nm and $W$ = 54, 54, 57 nm for the red, green, and blue rods respectively. The periodicities are $P_y$ = 330, 280, 120 nm respectively. (c) Optical bright field image of the plasmonic color primaries in reflection. The dark background next to the color pixels is the residual reflection of the glass substrate. The scale bar is 10 μm and the size of the pixel is 30 x 30 μm. (d) CIE chromaticity diagram with the location of the plasmonic color primaries under D65 standard illumination indicated by black stars. The white triangle represents the sRGB color space and the white circle represents the white color.

The color pixels are fabricated using electron beam lithography and dry etching of an evaporated silver film on an anti-reflection coated glass substrate (see methods section for details). The resulting nanorod structures are 45 nm tall with the length varying between $L$ = 63–143 nm and the width varying between $W$ = 54–57 nm depending on the represented color. The specific sizes of the nanorods to fulfil the resonance condition for the designed color were determined from numerical full wave simulations (see Supporting Information S2). All fabricated structures are coated with a 2 nm thick layer of alumina to prevent the silver from oxidizing[34]. To quantify the chromaticity and luminance of the surface, reflectance spectra are acquired on an inverted microscope using a 50x objective and a grating spectrometer (see Supporting Information S3 for a schematic of the experimental set-up). For the measurement, the pixels are illuminated through the objective with white light from a xenon arc lamp at normal incidence with linear polarization along the long axis of the nanorods. The reflected light is collected by the same objective. To include only the spectral contributions of the reflected light from the color pixel, a pinhole is inserted into an image plane formed in the detection path of the microscope. This serves the same purpose as a spectrometer slit and limits the area of collected light from the pixel to a circular area with a diameter of ~7 µm. The normalization to the power spectrum of the light source is performed by measuring the light reflected by a thick silver mirror from the same area. Additionally, optical bright field images in reflection are acquired with a 20x objective on a commercial microscope inspection system.

We design three primary color pixels to give red, green and blue color which span the plasmonic RGB color gamut. Figure 1b shows the measured reflectance spectra of these primary color pixels. The generated resonances are narrow with a full width at half maximum ranging between 55–82 nm, which puts $\Delta\lambda/\lambda$ to less than 16%. The nanorods of the color pixels are arranged with a periodicity $P_x$ of 426 nm, 355 nm and 284 nm fixing colors to red, green and blue respectively. Optical images in Fig. 1c illustrate the bright and vivid colors that are obtained in reflection with these color pixels. The pixels show a high color uniformity and no color variations are observed towards the edges of the pixels. To objectively classify the human perception of these colors, we calculate their chromaticity from the measured reflectance spectra. For this, the perception of a color of a given wavelength by the human eye needs to be considered. This perception is dominated by responsivity of the cone cells on the retina of the eye, which are described with color matching functions[35]. Taking these into account and a standard illumination D65, corresponding to an average midday light, the chromaticities $x$ and $y$ are obtained from the reflectance spectra (see Supporting Information Note 1 for the detailed calculation). For comparison, the colors are then located in the CIE 1931 color space chromaticity diagram. A large variation of chromaticity is obtained, resulting in an area of 55% of the sRGB color gamut that is covered with the generated plasmonic primary colors. These three primary color pixels span our plasmonic RGB color gamut and serve as a basis for our color mixing approach.

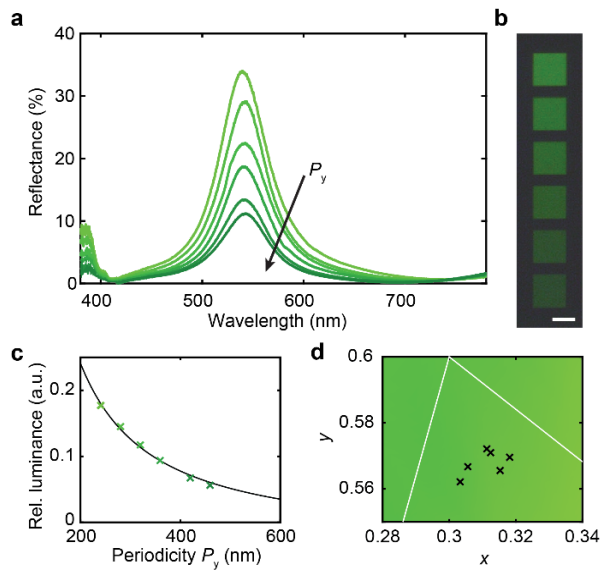

**Figure 2. | Adjustable color luminance of the green plasmonic primary color.** (a) Measured reflectance spectra of the green plasmonic primary color pixels with $P_y$ increasing from 240 nm to 460 nm. The black arrow indicates the direction of increasing $P_y$. The nanorod lengths and widths in all the pixels are $L$ = 102 nm and $W$ = 54 nm. (b) An optical bright field image of the green pixels shown in (a) arranged with successively increasing $P_y$ from top to bottom. The scale bar is 10 μm. (c) Relative luminance of the color pixels with increasing $P_y$. (d) Location of the color pixels in the green region of the CIE chromaticity diagram for varying $P_y$ under D65 standard illumination. The area of the chromaticity diagram in (d) is the one highlighted in Fig. 1d with dashed white lines.

Another equally important aspect of a color apart from its chromaticity is its luminance, i.e. how bright a color is perceived. The luminance of a color is obtained from the integral of the color matching functions, the measured reflectance and the illuminant spectral power distribution (see Supporting Information Note 1). While the chromaticity of single color pixels is set by the periodicity $P_x$, we specifically design our color pixels such that the color luminance is independently set by the orthogonal periodicity in *y* direction, $P_y$. Figure 2a shows reflectance spectra of a green color pixel with increasing lattice periodicity $P_y$ from 240 nm to 460 nm. The line shape of the reflectance spectra of the different pixels is close to identical, but with a decreasing maximum intensity for increasing $P_y$. With respect to the color, this corresponds to a decreasing color luminance at constant color chromaticity. Figure 2b shows the corresponding optical images of the pixels (measured in Fig. 2a) and illustrates the decrease of color luminance with increasing $P_y$. The measured relative luminance of these pixels is shown in Fig. 2c and exhibits a smooth decrease with increasing periodicity according to an inverse relation $\sim 1/P_y$. This dependence of the luminance on $P_y$ is particularly interesting, as it allows determining the periodicity $P_y$ to produce a given luminance. Furthermore, this decrease in luminance results in only negligible changes (less than ± 0.01) in the chromaticity of the color, as observed from the representation of the color chromaticity of the different pixels in the CIE diagram shown in Fig. 2d. Only in the limit of maximum luminance, i.e. when the periodicity $P_y$ becomes increasingly small, near-field interactions between adjacent nanorods slightly modify the color chromaticity. A similar analysis for adjusting the luminance of the red and blue color is shown in Supporting Information S4 and S5.

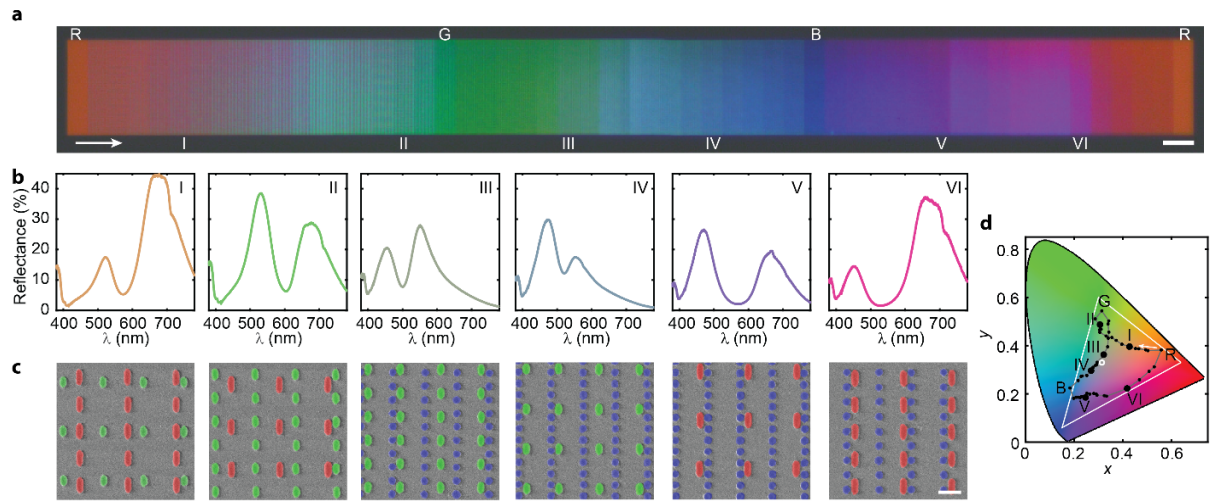

**Figure 3. | Plasmonic two-color mixing.** (a) Optical bright field image of a color gradient obtained by two-color mixing. In the first section nanorod arrays generating red (R) and green (G) are mixed, then arrays generating green and blue (B), and finally blue and red are gradually mixed. This defines the contour of accessible colors. The gradient is divided into 55 colors with the pixels of the primary colors consisting of only a single type of nanorod array marked with R, G and B. The respective nanorod lengths and widths are $L$ = 143, 102, 63 nm and $W$ = 54, 54, 57 nm for the red, green, and blue rods, respectively, and are the same for the entire gradient. The scale bar is 20 μm. (b) Measured reflectance spectra of six pixels at the marked positions I-VI of the gradient in (a). (c) False colored scanning electron micrographs of a section of the six pixels at the marked positions I-VI of the gradient in (a). The false coloring of the nanorods indicates the primary color that is generated by the respective nanorod arrays. The scale bar is 200 nm. (d) Measured colors along the color gradient in (a) represented with black dots in the CIE 1931 chromaticity diagram. The large black dots correspond to the chromaticities of the pixels I-VI and the corners marked R, G, B correspond to the chromaticities of the primary colors. The gray line connecting the black dots represents the contour of accessible colors. The white arrow represents the direction of travel along the gradient as marked also in (a).

Smooth transitions between multiple colors are obtained by spatially multiplexing the nanorod lattices of the different primary colors on a single pixel without increasing the pixel size, and adjusting the luminance of each color independently. The spatial multiplexing is performed by interleaving the nanorod lattices with their respective periodicities $P_{x1}$, $P_{y1}$ and $P_{x2}$, $P_{y2}$ in a geometrically non-conflicting manner. The superposition of the two periodic lattices leads to occasional overlap of two individual nanorods for non-multiple periodicities in the colors. This is avoided in a simple manner with the occasional shifting of one of the overlapping rods with respect to the other along the $x$ direction. Since this is only necessary for a few nanorods of a unit cell, the lattice is minimally distorted. Interleaving the lattices this way creates a multi-resonant dual color pixel where the chromaticity of the two colors is set by the respective $P_{x1}$ and $P_{x2}$ and the luminance of both colors can be freely and individually adjusted by $P_{y1}$ and $P_{y2}$ (see Supporting Information Note 2 for details).

An optical bright field image of a color gradient with smooth color transitions ranging from red to green, to blue and again to red color, which is obtained by the mixing of two colors through spatial multiplexing, is depicted in Fig. 3a. The color gradient is divided into 55 pixels and constituted of three sections of color mixing. In the first section, the red and green colors are mixed using a superposition of the nanorod lattices for the red and green primary color. In the second section green and blue are mixed, and finally blue and again red. The smooth transitions are obtained by varying the luminance of each color independently by adjusting their respective periodicities $P_{y1}$ and $P_{y2}$, while the periodicities setting the color chromaticity $P_{x1}$ and $P_{x2}$ are held constant. The measured reflectance spectra at different positions, marked (I – VI), along the color gradient and corresponding scanning electron microscope images of pixels are illustrated in Fig. 3b and c respectively (see Supporting Information S6 for the specific values of the periodicity $P_y$ for each color along the gradient). The measured reflectance shows that the smooth variation in color is attained by independently varying the color luminance of the respective primary colors. The spectral separation of the different resonant nanorod lattices results in a close to negligible crosstalk between different lattices that are interleaved. While near-field interactions between closely spaced scatterers do affect the resonance of individual nanorods (separations can be as close as 20 nm), the observed spectral shifts of less than ±15 nm (depending on the color) in the reflectance of the nanorod lattice are relatively small (see Supporting Information S6 for the measured reflectance spectra of every pixel along the gradient). As a result, color mixing of two primary colors with a close to independent adjustment of the individual color contributions is attained. This is in contrast to other color mixing schemes, which rely either on spatially separated pixels for each primary color[25] or on fixed mixing ratios[16,19].

Figure 3d summarizes the chromaticity values of the color gradient of Fig. 3a in the chromaticity diagram. The binary mixing of the three color primaries results in a smooth color variation along the three depicted solid black lines that span the plasmonic RGB color gamut (pRGB). The pixels highlighted in Fig. 3b and c are marked with black dots. The plasmonic color gamut spanned by our color mixing method covers 39% of the sRGB color gamut. The pRGB gamut forms a distinct figure-eight-shaped area, as opposed to the triangle that is characteristic of linear color mixing. The curved lines connecting the primary colors as well as the crossing of the lines connecting green and red, and green and blue, are due to various nonlinearities in the mixing process such as small spectral shifts due to near-field interaction or chromaticity variation with $P_y$. Using an additional degree of freedom, namely the size of the nanorods, can be a direct way to correct for such nonlinearities, and to achieve a larger coverage of the CIE color gamut.

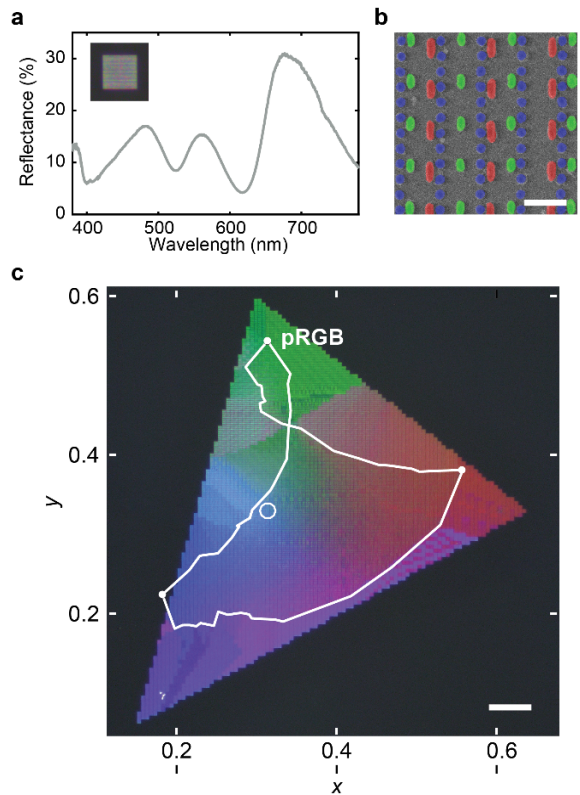

**Figure 4. | The triplet plasmonic pixel and the plasmonic RGB color gamut.** (a) Measured reflectance spectrum of a triplet color pixel producing the color white. The inset shows an optical bright field image of the pixel. (b) False colored scanning electron micrograph image of the white pixel. The false coloring of the nanorods indicates the primary color that is generated by the respective nanorod arrays. The scale bar is 300 nm. (c) Optical bright field image of the plasmonic RGB color space using three-color mixing. The colors contained within the area outlined by the solid white lines are expected to be reproduced correctly. The scale bar is 50 μm. The nanorod lengths and widths in (a), (b) and (c) are $L$ = 143, 102, 63 nm and $W$ = 54, 54, 57 nm for the red, green, and blue rods, respectively.

Another essential part of a color-mixing scheme is the representation of a white color. Adapting our method to three primary colors enables the generation of a white pixel by using the suitable combination of the periodicities $P_{y1}$, $P_{y2}$ and $P_{y3}$ of three interleaved nanorod lattices. A reflectance spectrum and an optical bright field image of a white pixel created in this manner by three-color mixing are shown in Figure 4a. The reflection spectrum shows the three distinct peaks of the red, green and blue primary colors necessary to create white. The relative intensities of the three peaks required to obtain a white color are adjusted to the perception of the three colors by the human eye, and therefore, have different contributing intensities. Green colors are typically perceived brighter than red or blue color tone. Therefore, an equal mixture of perceived red, green and blue, as is required for creating white color, results in a higher intensity of red and blue as compared to the green. The three interleaved nanorod lattices to generate white are illustrated in the SEM image in Fig. 4b. Here, again the overlapping of nanorods is avoided with a similar method as in the two-color mixing case. The demonstrated white pixel exhibits chromaticity values of $x$ = 0.308 and $y$ = 0.333 which is within less than 1% in chromaticity from the ideal white (0.3127, 0.329) for D65 standard illumination.

Independently mixing three primary colors allows for reproducing colors that continuously cover the plasmonic color gamut illustrated in Fig. 3d. The periodicity triplet ($P_{y1}$, $P_{y2}$, $P_{y3}$) of each color in this gamut is determined from extrapolating the measured chromaticity values and periodicities for the two-color mixing case, assuming independent mixing of the colors. This periodicity triplet is analogous to the RGB triplet that gives the respective intensities of the three primary colors in conventional RGB color mixing. Correspondingly, an arbitrary fine variation of these triplet values of this plasmonic RGB pixel enables the realization of a countless number of color hues and tones, and seamless gradients between different colors.

To highlight the effectiveness of our plasmonic RGB pixels to realize arbitrary color tones with fine color variations, an image of the entire plasmonic RGB color gamut is fabricated. Figure 4c shows an optical bright field image of this plasmonic RGB color gamut realized using our multi-color triplet plasmonic pixels. Here, the solid white lines represent the border of the plasmonic color gamut we discussed previously (Fig. 3d). Colors with chromaticity values outside of the pRGB gamut are not represented correctly by our triplet plasmonic pixels. The values of the periodicity triplets for each pixel are shown in Supporting Information S6. Figure 4c shows that a seamless color variation is indeed obtained in experiment, and as opposed to only specific colors in the CIE 1931 color diagram, a continuous area is smoothly covered. With a pixel size of 4.26 x 4.26 μm, the representation of the color gamut includes 2456 colors, each with a slightly different color chromaticity and periodicity triplet. This number can easily be increased by fabricating the gamut on a larger area, leading to even finer transitions between chromaticity values.

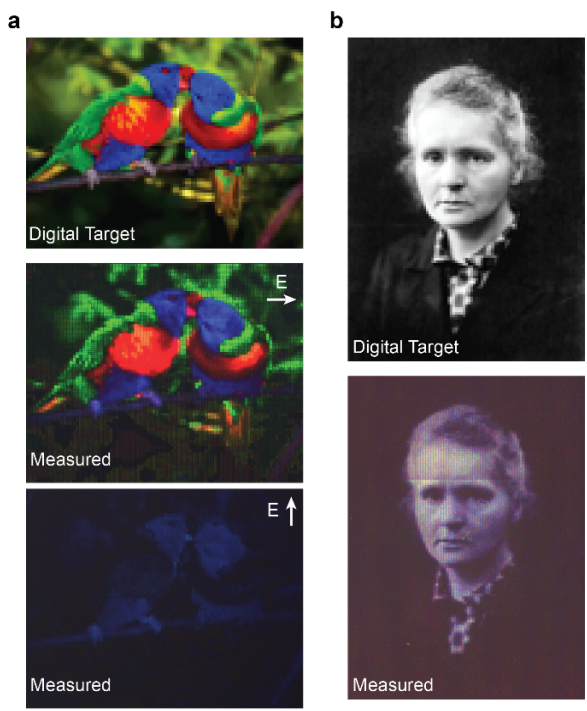

**Figure 5. | Photorealistic color printing with pRGB coloring.** (a) Digital image of two colorful parrots and its corresponding measured optical bright field image generated using pRGB color mixing acquired with linearly polarized light along the nanorod long axis (middle) and along short axis (bottom). The image

shows smooth transitions between the different color hues and tones as enabled by our color mixing method. The fabricated image is 307 x 400 µm in size with pixels of 4.26 x 4.26 µm. (b) Digital grayscale portrait of Marie Skłodowska Curie and its measured optical bright field image. The fabricated image is 533 x 388 µm in size with pixels of 4.26 x 4.26 µm. The nanorod lengths and widths in (a) and (b) are L = 143, 102, 63 nm and W = 54, 54, 57 nm for the red, green, and blue rods, respectively.

As a demonstration of our plasmonic coloring method, we reproduced color and grayscale photographs. Figure 5a shows the digital target image and an optical bright field image of the fabricated photograph of two colorful parrots realized using plasmonic RGB color mixing. The smooth transitions in color and brightness in the feathers of the parrots and in the background are nicely represented using our coloring approach. In addition, illuminating the fabricated image with light polarized perpendicular to the long axis of the nanorods, results in close to negligible reflected intensity. This polarization sensitive behavior could be used for introducing active tunable colors[13,15] or for anti-counterfeiting applications[7]. Figure 5b shows a grayscale target image and the corresponding optical bright field image of the fabricated surface portraying Marie Skłodowska Curie. Similar to before, here the smooth grayscale transitions on the cheeks and neck are nicely represented in our fabricated image as in a realistic photograph. Comparing both images to their target, we observe that colors and color transitions are nicely represented, but especially colors with very low brightness are difficult to reproduce correctly. Remarkably, our method allows the reproduction of grayscale and color images from identical building blocks, as opposed to previous work, where these could not be represented by the same structure on the same surface[4]. In contrast to previous work that relied on color generation with only a single resonance mode, our method allows the reproduction of images with smooth transitions between different color hues and tones.

The performance of our color pixels and our color mixing approach can be further advanced in future studies depending on applicational interest. In addition to focusing only on color mixing by adjusting the periodicity $P_y$, the lengths of the nanorods can be used as additional degrees of freedom. With this or by choosing different primary colors, and thus periodicities $P_y$, such as yellow or cyan, or by adding more colors to create an effective four- or five-color mixing, an even larger color gamut can be obtained. Furthermore, the use of scatterers of different shape[14] or materials[16,19,20,36] for obtaining high-quality resonances may also result in an increase in the vividness and brightness of the colors in the presented plasmonic color gamut. With respect to the fabrication, the process used in the present work is already reasonably fast. For example, the image shown in Fig. 5a was written in under 5 min with an electron beam lithography exposure optimized for minimal number of beam shots per individual nanorod. However, for large scale applications we envision the realization of these structures with nanoimprint lithography[37].

Our method of color generation and color mixing has potential application in a host of technologies, ranging from anti-counterfeiting security[38] to multiband optical filters[39]. Other applications are in aesthetic color coatings for solar cells[40] or color-sensitive optoelectronic sensors[41], where partial transparency is required. Additionally, introducing a dynamic tuning of the structural resonances[42–45] would allow the realization of passive transparent displays for consumer electronics. Furthermore, multi-resonant arrays such as the ones proposed in this work may enable new lattice based phenomena such as multi-color lasing or white light lasing from a periodic lattice[46,47].

## Summary

We have presented a facile method for structurally generating and mixing vivid colors additively based on interleaved rectangular lattices of metallic nanoparticles. This coloring approach offers an unprecedented individual control over the chromaticity and luminance of the generated colors. We generate vivid primary colors, red, green and blue, and superimpose these on a single multi-resonant triplet pixel where the contribution of each primary color can be individually adjusted. Based on these, we demonstrate an original, additive plasmonic RGB color mixing approach that allows representing countless numbers of colors within a color gamut that continuously spans 39% of the sRGB color gamut. The demonstrated level of control over color chromaticity and brightness, the continuous coverage of a color gamut and the photorealistic reproduction of images represent a large step forward compared to the state of the art, where the luminance of the colors cannot be easily adjusted[9,10,21,26] or other color mixing approaches where the number of colors are limited[16,19,25]. Furthermore, our method of using a single multi-resonant triplet color pixel increases the resolution by a factor of 7 and greatly increases color uniformity, as compared to a recently demonstrated CMYK color mixing approach[25]. As a result, we demonstrate photorealistic reproductions of color and grayscale images with smooth transitions between different colors and different lightness of colors.

## Methods

**Measurement** The fabricated samples are characterized on a home-built inverted microscope (schematically shown in Supporting Information S3) to measure the reflectance of the structures. White light from a Xenon lamp impinges on the sample with linear polarization along the nanorod long axis and light reflected from the color pixels is collected using an air objective (NA=0.5). The collected light is sent to a grating spectrometer. Optical bright field images are acquired on a Leica DM8000M microscope using a 20x objective and white light from an LED polarized along the long axis of the nanorods. For the bright field images, the white balance is performed on a silver mirror.

**Fabrication** The surfaces are fabricated on borosilicate glass substrates (*n* = 1.52) with a four layer anti-reflection coating (Edmund Optics VIS0). To remove organic residues from the surface, the substrates are cleaned in an ultrasonic bath in water, acetone and IPA each for 15 min, dried using a $N_2$ gun and subsequently cleaned using oxygen plasma. A 45 nm thick layer of silver is evaporated directly onto the glass using an electron beam evaporator. A 1 nm titanium adhesion layer is used to avoid delamination of the structures[48]. In a next step, the nanorods are written in a spin coated HSQ resist layer by standard electron beam lithography. Before spin coating the substrate is immersed for 30 s in Surpass 3000 for surface promotion, rinsed in water for 20 s and dried using a nitrogen gun. To improve the adhesion of the resist to the bare silver, a 5 nm thick layer of $SiO_2$ is sputtered onto the metal coated substrate, which also prevented the oxidation of the silver film during the processing. To remove the unexposed resist, the substrate is immersed into an aqueous developer solution (1% NaCl, 4% NaOH). The nanoantenna pattern is transferred to the silver with argon ion sputter etching. For the removal of residual resist, the samples were then immersed in buffered hydrofluoric acid (1:7) for 5 s, subsequently rinsed in DI water, and cleaned in acetone and IPA. To protect the silver nanostructures from oxidizing, a 2 nm thick $Al_2O_3$ layer was deposited using a low temperature atomic layer deposition process at 50°C.

**Simulations** The numerical modelling of the nanostructures is done using a finite difference time domain (FDTD) method. Simulations are performed with a commercially available FDTD software (Lumerical FDTD Solutions). Tabulated values from Palik[49] are used for the dielectric function of silver and $Al_2O_3$, and a constant refractive index of 1.52 for the borosilicate glass. The simulations are carried out with a coherent plane wave source polarized along the nanorod long axis. Perfectly matched layer boundary conditions are applied on all sides of the computational domain unless otherwise noted. A smallest mesh-refinement of 1 nm is used around the silver nanorods. The anti-reflection coating is modelled with a four layer coating of refractive indices $n$ = 1.385, 2.15, 1.62 and 1.385 and thicknesses $h$ = 90, 107, 50 and 150 nm from top to bottom. These values are determined by matching the reflection measured by the supplier by using a multilayer transfer matrix method calculation.

**Acknowledgements**

We would like to thank Dr. Antonis Olziersky for his help with electron beam lithography.

**Contributions**

C.U.H. and H.E. designed the experiments and measurement setups. C.U.H. designed and fabricated the devices, and performed theoretical and experimental analysis. M.D. contributed to the design of the multi-color pixels. H.E., G.S. and D.P. supervised all aspects of the project. C.U.H., G. S., H. E. and D.P. wrote the manuscript. All authors commented on the manuscript.

**Competing interests**

The authors declare no competing interests.

**Correspondence and requests for materials** should be addressed to H.E. or D.P.

**Data availability**

The data that support the plots within this paper and other findings of this study are available from the corresponding authors upon reasonable request.

**Supporting Information:**

**A plasmonic painter's method of color mixing for a continuous RGB palette**

**Authors**

Claudio U. Hail[1], Gabriel Schnoering[1], Mehdi Damak[1], Dimos Poulikakos[1]*, Hadi Eghlidi[1]*

**Affiliations**

[1] Laboratory of Thermodynamics in Emerging Technologies, ETH Zürich, Sonneggstrasse 3, CH-8092 Zürich, Switzerland.

* Correspondence and requests for materials should be addressed to D.P. (email: dpoulikakos@ethz.ch) or H.E. (email: eghlidim@ethz.ch).

## Supplementary Figures

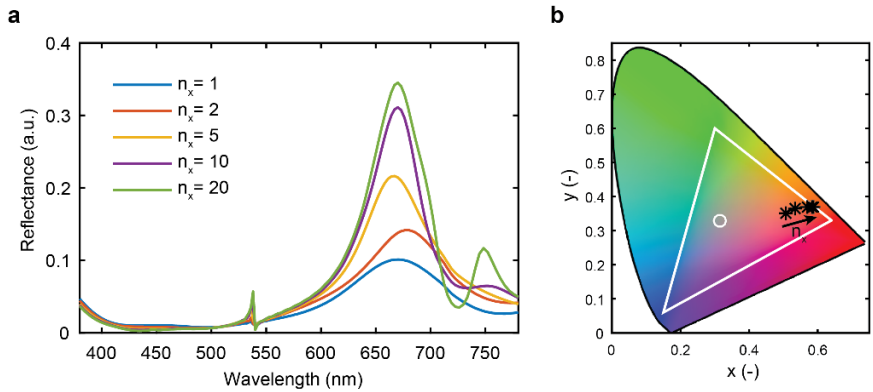

**Supporting Figure S1 | Numerical analysis of the effect of nanorod array periodicity.** (a) Numerically simulated reflectance of a silver nanorod array with height $H$ = 45 nm, width $W$ = 54 nm, length $L$ = 135 nm and varying number of periods along the $x$ direction, $n_x$ (see Fig. 1a). Periodic boundary conditions were applied along the $y$ direction of the computational domain. (b) Locations of the simulated reflected colors in the CIE chromaticity diagram with increasing number of periods $n_x$. By increasing the number of periods, a close to 4-fold increase in maximum intensity and a spectral narrowing is obtained, resulting in more vivid colors as clearly seen in panel (b).

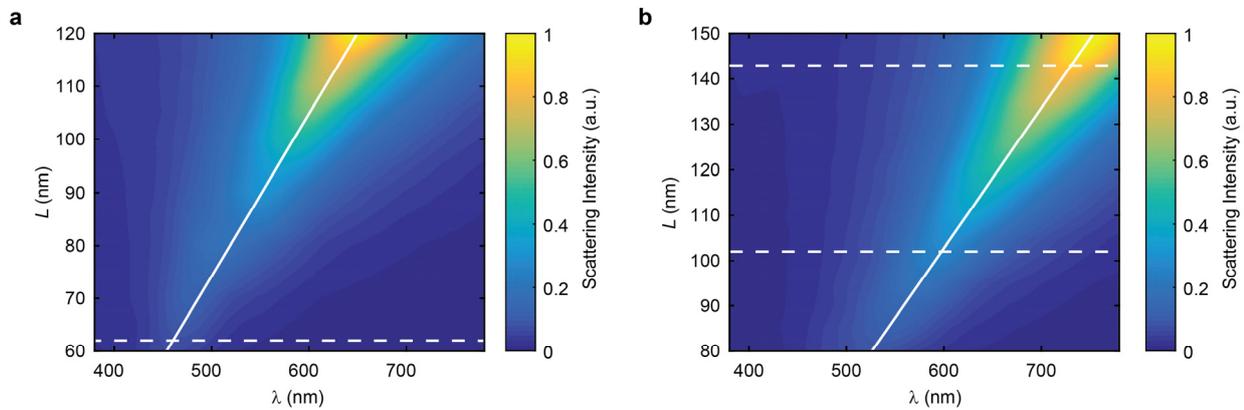

**Supporting Figure S2 | Numerical analysis of scattering from single silver nanorods.** (a) Numerically simulated backscattered intensity of a silver nanorod with height $H$ = 45 nm, width $W$ = 57 nm and varying length $L$. The white dashed line marks the length of the nanorod generating the blue primary color. (b) Numerically simulated backscattered intensity of a silver nanorod with height $H$ = 45 nm, width $W$ = 54 nm and varying length $L$. The white dashed lines mark the lengths of the nanorods generating the green and red primary colors respectively. The solid white line represents a linear fit to the maximum scattering intensity for varying rod lengths.

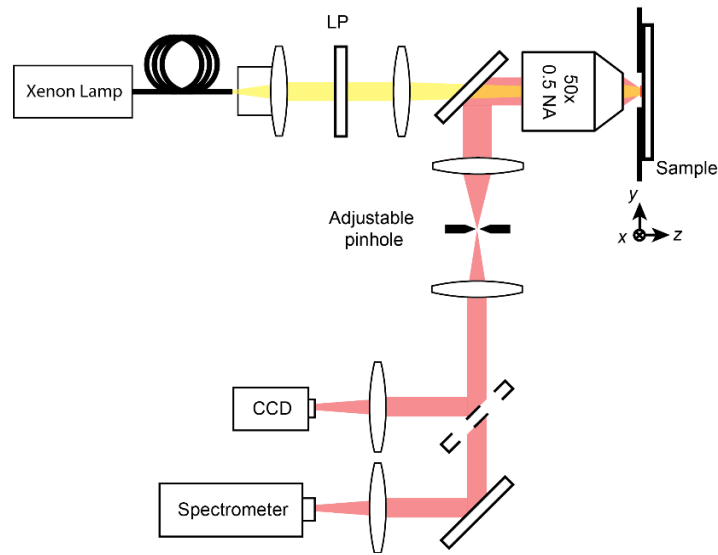

**Supporting Figure S3 | Experimental setup for reflectance measurements.** The fabricated samples are illuminated in Köhler illumination with white light from a Xenon Lamp (Newport) passing through a beam splitter and an imaging objective (50x, 0.5 NA). The light reflected from the sample is collected by the same objective and reflected by the beam splitter. A linear polarizer (LP) is used to set the polarization of the white light along the long axis of the fabricated nanorods on the sample. In the detection path, an image plane is formed and an adjustable pinhole is inserted to limit the area of the substrate to acquire a spectrum from. This area is maintained constant for all measurements. A charge coupled device (CCD, PCO pixelfly) is used to locate the color pixels on the substrate. For acquiring a reflectance spectrum, the mirror shown in a dashed outline, is removed and light is directed to the grating spectrometer (Princeton).

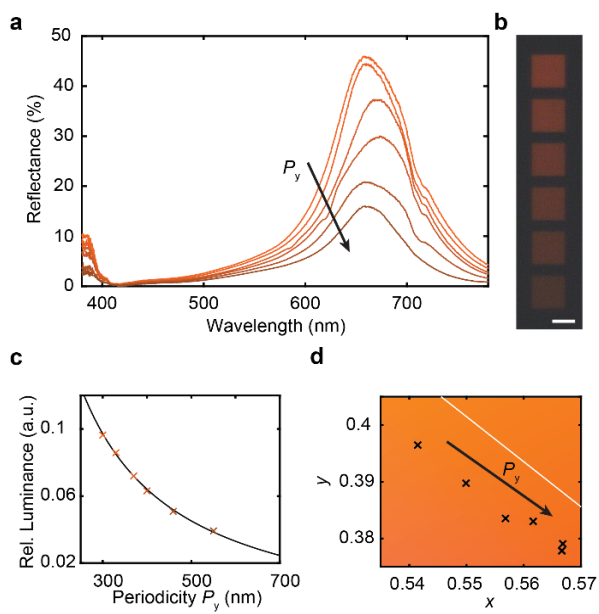

**Supporting Figure S4 | Adjustable color luminance of the red plasmonic primary color.** (a) Measured reflectance spectra of the red plasmonic primary color pixels with $P_y$ increasing from 300 nm to 550 nm. The black arrow indicates the direction of increasing $P_y$. (b) Optical bright field image of the red pixels shown in (a) with increasing $P_y$ from top to bottom. The scale bar is 10 μm. (c) Relative luminance of the color pixels with increasing $P_y$. (d) Location of the color pixels in the CIE Chromaticity diagram for varying $P_y$.

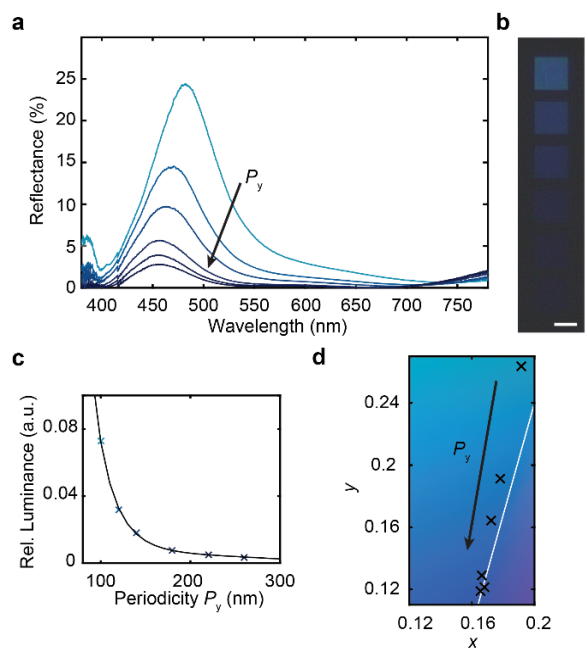

**Supporting Figure S5 | Adjustable color luminance of the blue plasmonic primary color.** (a) Measured reflectance spectra of the blue plasmonic primary color pixels with $P_y$ increasing from 100 nm to 260 nm. The black arrow indicates the direction of increasing $P_y$. (b) Optical bright field image of the blue pixels shown in (a) with increasing $P_y$ from top to bottom. The scale bar is 10 μm. (c) Relative luminance of the color pixels with increasing $P_y$. (d) Location of the color pixels in the CIE Chromaticity diagram for varying $P_y$.

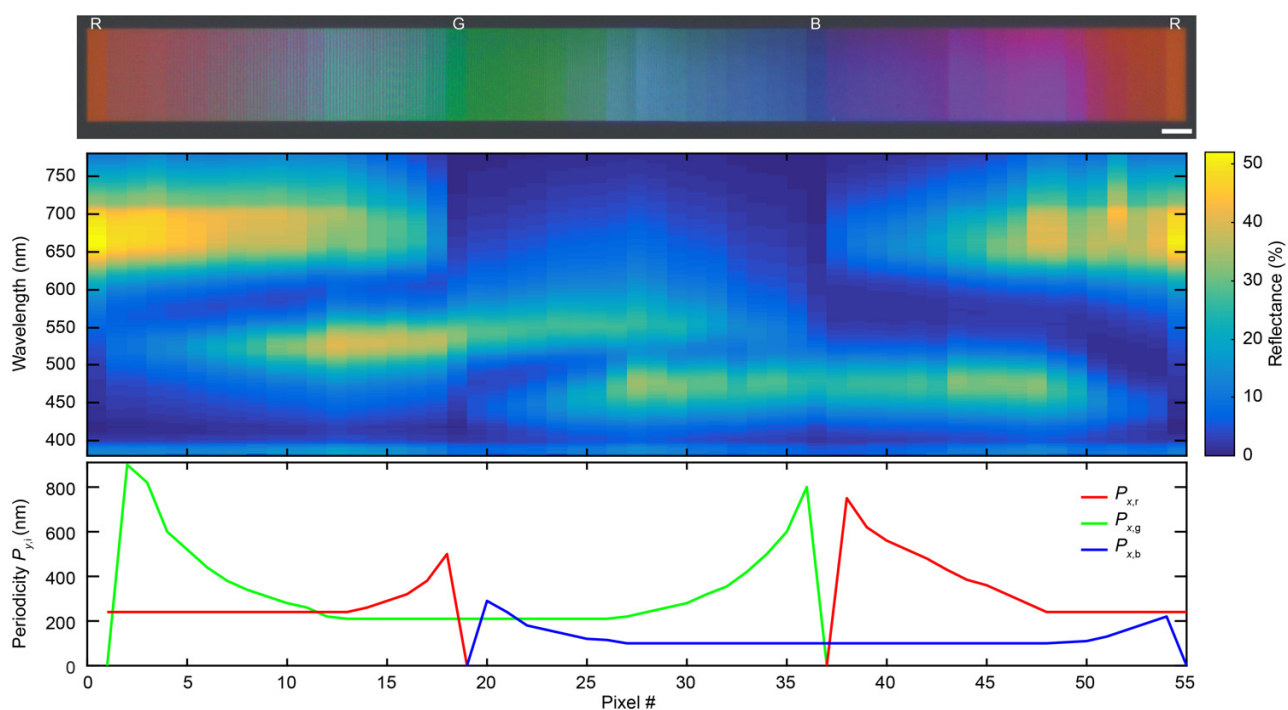

**Supporting Figure S6 | Color gradient reflectance and periodicities.** An optical bright field image of the color gradient shown in Fig. 3a and the corresponding measured reflectance spectra at each of the reproduced 55 pixels. The variation of the periodicity $P_y$ used for generating the color gradient is illustrated in the bottom panel. The periodicity of the rectangular arrays are shown in the respective primary color that is generated by the nanorod array. The x-axis of the reflectance spectra and periodicity plot align with the corresponding pixels in the optical image.

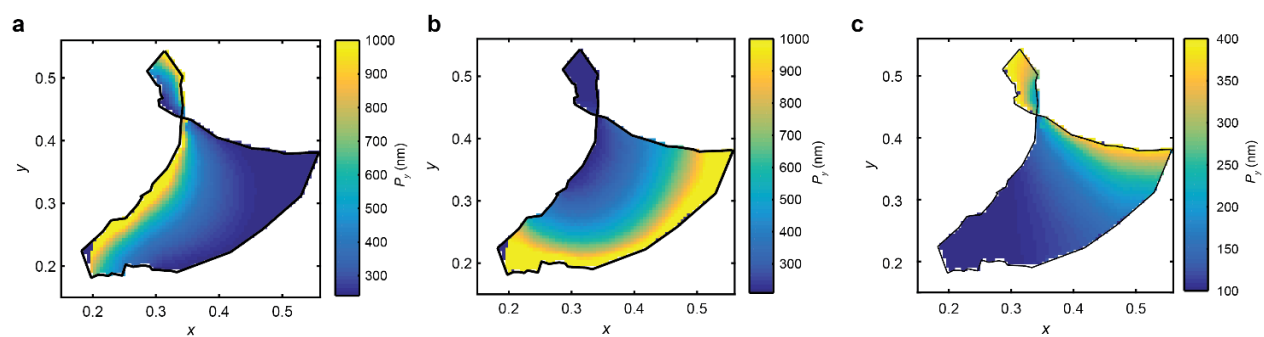

**Supporting Figure S7 | Periodicity of the pixels in the pRGB color gamut.** Periodicities in the y direction of the nanorod arrays for the red (a), green (b), and blue (c) primary color for representing the pRGB color gamut.

## Suppporting Note 1: Color chromaticity and luminance calculation

The CIE coordinates of a color pixel is calculated using the following integrals

$$X = \int_{380}^{780} I(\lambda)R(\lambda)\bar{x}(\lambda)d\lambda \qquad (1.1)$$

$$Y = \int_{380}^{780} I(\lambda)R(\lambda)\bar{y}(\lambda)d\lambda \qquad (1.2)$$

$$Z = \int_{380}^{780} I(\lambda)R(\lambda)\bar{z}(\lambda)d\lambda, \qquad (1.3)$$

where $R(\lambda)$ corresponds to the measured reflectance, $I(\lambda)$ is the spectral power distribution of the illuminant and $\bar{x}(\lambda)$, $\bar{y}(\lambda)$ and $\bar{z}(\lambda)$ are the respective color matching functions. The integrals are evaluated over the entire visible wavelength range from 380 nm to 780 nm. Figure S8 shows the spectral dependence of these color-matching functions. From the coordinates *X, Y* and *Z*, the normalized chromaticities *x* and *y* are then obtained by

$$x = \frac{X}{X+Y+Z} \text{ and } y = \frac{Y}{X+Y+Z}, \qquad (1.4)$$

and can be illustrated in the CIE 1931 color diagram. The luminance of a color corresponds directly to the *Y* coordinate. As observed from Fig. S8, the color-matching function $\bar{y}(\lambda)$ has a maximum value at green wavelengths. As a consequence, green colors are seen brighter by the human eye as compared to blue or red.

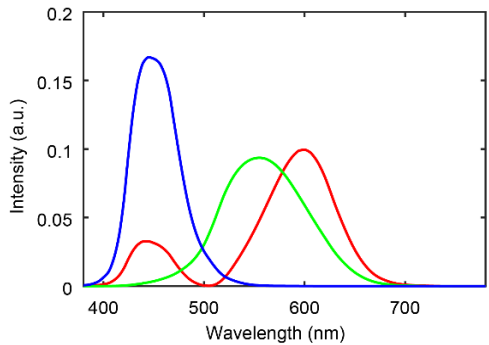

**Supporting Figure S8 | Color matching functions.** The spectral dependence of the color-matching functions of the human eye, $\bar{x}(\lambda)$, $\bar{y}(\lambda)$ and $\bar{z}(\lambda)$ [1].

## Supporting Note 2: Multiplexing of rectangular nanorod arrays

An essential aspect of our color mixing approach is the spatial multiplexing of two or three nanorod arrays with different periodicities in the $x$ and $y$ direction. This multiplexing is achieved by superimposing two or three arrays with fixed periodicities in $x$ direction, in our case $P_{x,r}$ = 426 nm, $P_{x,g}$ = 355 nm and $P_{x,b}$ = 284 nm for the red, green and blue color arrays respectively. For tuning the color luminance smoothly, the periodicities in $y$ direction of all the three primary color arrays need to be continuously adjustable. This requirements can be met if one dimensional periodic arrays with the fixed periodicities $P_x$ can be superimposed without any overlap.

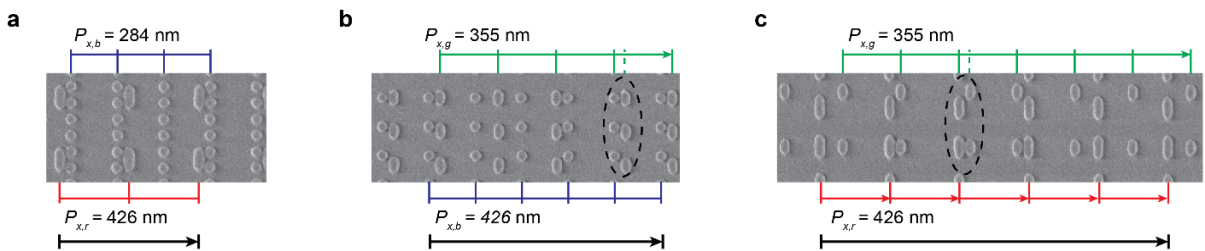

**Supporting Figure S9 | Two color mixing arrangements.** (a) SEM of a unit-cell for red and blue color mixing. (b) SEM of a unit-cell for green and blue color mixing. (c) SEM of a unit-cell for green and red color mixing. The red, green and blue lines highlight the periodicities of the respective nanorod lattice. The black arrow highlights the smallest unit-cell size. The dashed black ellipses highlight the location where an overlap is avoided by displacing the green nanorods along the $x$ direction with respect to the others.

For two color mixing, the three cases of mixing red and blue, green and blue, and red and green can be separated. Figure S9 shows a scanning electron micrograph of a unit cell of each of these cases of two color mixing. For mixing red and blue, the two arrays can be conveniently superimposed when offset by 71 nm. This is due to the relatively large greatest common divisor of $P_{x,r}$ and $P_{x,b}$. For mixing green and blue, and red and green, more attention is required due to the reduced largest common divisor of the respective $P_x$. Offsetting the green array with respect to the blue one by 71 nm, results in an overlap of the green and blue rods in the area highlighted by the black dashed ellipse in Fig S9b. As observed in the scanning electron micrograph, this is avoided by displacing the green rod by 71 nm, thus by 2/5 π in phase of the periodic structure. In a similar manner a collision of the red and green rod is avoided when mixing these arrays together as shown in Fig S9c.

For three color mixing a similar approach is followed to ensure no overlaps between the different arrays. Figure S10 shows a unit cell of a surface where the three lattices are superimposed to create three color mixing. The unit cell size is 4.26 μm in the *x* direction and freely adjustable in the *y* direction. The four locations of overlap between the structures are marked with the black dashed ellipses. In these cases, the green rod is shifted by 71 nm in positive or negative *x* direction to avoid the overlap. By choosing periodicities with even a larger greatest common divisor, overlapping between the different arrays could also be avoided.

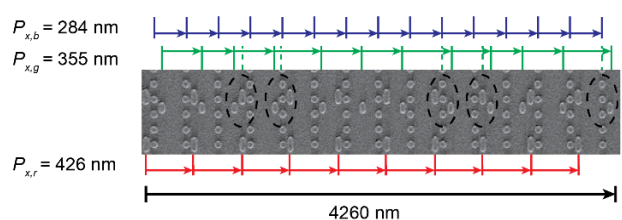

$P_{x,b}$ = 284 nm
$P_{x,g}$ = 355 nm
$P_{x,r}$ = 426 nm
4260 nm

**Supporting Figure S10 | Three color mixing arrangement.** (a) SEM of a unit-cell for three color mixing. The red, green and blue lines highlight the periodicities of the respective nanorod lattice. The black arrow highlights the smallest unit-cell size. The dashed black ellipses highlight the location where an overlap is avoided by displacing the green nanorods along the *x* direction with respect to the others.